\documentclass[review,authoryear,10pt]{elsarticle}

\usepackage{amssymb}

 \usepackage{lineno}
 \usepackage{amsfonts,amssymb,cite}
 \usepackage[latin1]{inputenc}
 \usepackage{enumerate}
 \usepackage{amssymb}
 \catcode`\@=11 \@addtoreset{equation}{section}
 
 \catcode`\@=12
 \usepackage{latexsym}
 \usepackage{multirow}
 \usepackage{textcomp}
 \usepackage{amsmath}
 \usepackage{amsfonts}
 \usepackage{dsfont}
 \usepackage{amssymb}
 \usepackage{mathrsfs}
 \usepackage{dsfont}
 \usepackage{graphicx}
 \usepackage{bbm}

 \graphicspath{ {images/} }
 \DeclareGraphicsExtensions{.png,.pdf}

 \newcommand{\be}{\begin{equation}}
 \newcommand{\ee}{\end{equation}}

 \newtheorem{Theorem}{Theorem}[section]

 \usepackage{multirow}

\journal{Statistics and Computing}

\begin{document}

\begin{frontmatter}



\title{Variable selection with Group LASSO approach : \\ Application to Cox regression with frailty model }





\author[Affil1]{Jean Claude Utazirubanda}
\author[Affil2]{Tomas Leon}
\author[Affil1]{Papa Ngom \corref{cor1}}
\address[Affil1]{LMA,Universit\'{e} Cheikh Anta Diop, Dakar, Senegal}
\address[Affil2]{School of Public Health, University of California, Berkeley, USA}
\cortext[cor1]{Corresponding Author: Papa Ngom; Email: papa.ngom@ucad.edu.sn }

\begin{abstract}
In analysis of survival outcomes supplemented with both clinical information and high-dimensional gene expression data, use of the traditional Cox proportional hazards model fails to meet some emerging needs in biomedical research. First, the number of covariates is generally much larger the sample size. 
Secondly, predicting an outcome based on individual gene expression is inadequate because multiple biological processes and functional pathways regulate phenotypic expression. Another challenge is that the Cox model assumes that populations are homogenous, implying that all individuals have the same risk of death, which is rarely true due to unmeasured risk factors among populations. In this paper we propose group LASSO with gamma-distributed frailty for variable selection in Cox regression by extending previous scholarship to account for heterogeneity among group structures related to exposure and susceptibility. The consistency property of the proposed method is established. This method is appropriate for addressing a wide variety of research questions from genetics to air pollution. Simulated analysis shows promising performance by group LASSO compared with other methods, including group SCAD and group MCP. Future research directions include expanding the use of frailty with adaptive group LASSO and sparse group LASSO methods. 
\end{abstract}

\end{frontmatter}

\pagebreak









\section{Introduction}
\label{intro}
Survival analysis models the time it takes for death and other long-term events to occur, focusing on the distribution of survival times. Survival modeling examines the relationship between survival and one or more predictors, usually called $covariates$ in the survival-analysis literature. The standard modeled event is $death$, from which the name $survival$ $ analysis$ and much of its terminology derives, but the scope and applications of survival analysis are much broader. Similar methods are used in other disciplines with different outcomes of interest: operating time of a machine to measure $reliability$, $event$-$history$ $analysis$ of marriage, divorce, and unemployment in sociology, and duration of contracts in actuarial sciences (survival time $T$ from the execution until the cancellation or completion of a contract).

The semi-parametric approach is one of three approaches found in survival analysis. It is an intermediate method between the parametric and non-parametric approaches. In the semi-parametric approach, the real probability distributions of observations are assumed to belong to a class of laws dependent upon parameters, while other parts are written as non-parametric functions. This approach is commonly used in survival data analysis (Cox $ 1972$; Cox \& Oakes $1984$). 

By using the Cox regression model, we specifically aim to model the impact of predictors on the hazard function, which characterizes for an individual $ j $ the probability of dying or experiencing a particular outcome within a short interval of time provided the individual has survived or not experienced the outcome previously. It is useful for identifying the risk factors of a disease, comparing treatments, and estimating the probability of occurrence of an event such as death or relapse in a given identified individual with a vector of explanatory variables. Many extended versions of the Cox regression model have been implemented to take into account clustered data or groups within which the failure times may be correlated (Martinussen \& Scheike $2006$). These groups may represent such distinct entities as members of the same family, patients in the same hospital, or organs within an individual. These groups may also represent repeated timed observations in the same individual, including recurring symptoms of certain diseases or multiple relapses. 
Grouping structures arise naturally in many statistical modeling problems. As addressed by Ma et al., complex diseases such as cancer are often caused by mutations in pathways involving multiple genes; therefore, it would be preferable to select groups of related genes together rather than individual genes separately if they operate on the same causal pathway($2007$). 
In linear regression, many variable selection techniques have traditionally been used. Three examples are best subset and forward and backward stepwise selection, which produce a sparse model. Best subset regression finds for each $k\in \lbrace 1,..., p\rbrace$ the subsets of size $k$ that gives the smallest residual sum of squares. The question of how to choose $k$ involves the trade-off between bias and variance, along with the more subjective desire for parsimony. There are a number of criteria that one may use; typically, we choose the smallest model that minimizes an estimate of the expected prediction error.

However, this technique is often unsatisfactory for two reasons: 1) the number of "all possible subsets" grows exponentially with the number of predictors ($p$), so when the number of predictors ($ p$) is large, searching all possible subsets is computationally intensive and inefficient; 2) subset selection is discontinuous, implying that an infinitesimally small change in the data can result in completely different estimates. This causes the subset selection method to be unstable and highly variable, especially in higher dimensions (Breiman $1995$; Fani \& Li 2001). 

Rather than search through all possible subsets (which becomes infeasible for $p$ much larger than $40$), we can seek a guided path through them. $Forward-stepwise$ $selection$ starts with the intercept and then sequentially adds into the model the predictor that most improves the fit. Forward-stepwise selection is a $greedy$ $algorithm$, producing a nested sequence of models. In this sense it might seem suboptimal compared to best subset selection, but there are a few reasons why it might be preferred. First, computationally; for large $p$ we cannot compute the best subset sequence, but we can always compute the forward-stepwise sequence (even when the number of predictors $p$ is greater than the sample size $n$). 
Second, statistically; a price is paid in variance for selecting the best subset of each size; forward-stepwise selection is a more constrained search and will have lower variance but perhaps more bias. 
$Backward-stepwise$ $selection$ starts with the full model and sequentially deletes the predictor that has the least impact on the fit. The candidate variable for dropping is the one with the smallest Z-score. Backward selection can only be used when the sample size $n$ is greater than the number of predictors $p$, while forward stepwise can always be used (Hastie et al. 2009). While useful in many contexts, stepwise techniques (forward and backward) for variable selection are still unsatisfactory in certain situations (Greenland 2008).

Penalized regression techniques have been proposed to accomplish the same goals as the best subset selection and forward- and backward- stepwise selection but in a more stable, continuous, and computationally efficient fashion. These techniques include a $L_{1}$ absolute value "Least Absolute Shrinkage and Selection Operator" ("LASSO") penalty (Tibshirani $1996$, $1997$), and a $L_{2}$ quadratic ("ridge") penalty (Hoerl \& Kennard $1970$; Le Cessie \& van Houwelingen $1992$; Verweij \& Van Houwelingen $1994$). $L_{1}$ and $L_{2}$ penalized estimation methods shrink the estimates of the regression coefficients towards zero relative to the maximum likelihood estimates. The purpose of this shrinkage is to prevent overfitting due to either collinearity of the covariates or high dimensionality. Although both methods are shrinkage oriented, the effects of $L_{1}$ and $L_{2}$ penalization are quite different in practice. Applying a $L_{2}$ penalty tends to result in all small but non-zero regression coefficients. As a continuous shrinkage method, if there is high correlation between predictors, ridge regression achieves better predictive performance through a bias-variance trade-off that favors ridge over LASSO (Tibshirani $ 1996 $). However, ridge regression cannot produce a parsimonious model, as it produces coefficient values for each of the predictor variables. Applying a $L_{1}$ penalty tends to result in many regression coefficients shrunk exactly to zero and a few other regression coefficients with comparatively little shrinkage. Consequently, LASSO has become more popular due to its sparse output. 

The $L_{1}$ penalty has been applied to other models including Cox regression ( Tibshirani $1997$) and logistic regression ( Lokhorst $1999$; Roth $2004$; Genkin et  al. $2007$). Even though LASSO has been successfully utilized in many situations, its popularity and applications are still limited.
In the $p>n$ case, LASSO selects at most $n$ variables before it saturates because of the nature of the convex optimization problem. Moreover, LASSO is not well defined unless the bound on the $L_{1}$ norm of the coefficients is smaller than a certain value (Zou \& Hastie $2005$).
When predictors are categorical, the LASSO solution is not satisfactory, as it only selects individual dummy variables instead of whole factors and depends on how the dummy variables are coded (Meier et al. $2008$). This process results in models that are dependent upon how categories are defined and may produce findings that are artifacts of this arbitrary nature and use of breakpoints.
The group LASSO method is an extension of this popular model selection and shrinkage estimation $L_{1}$ penalty technique to address the problem of variable selection in high dimensions $(i.e.,$ the number of regressors $p$ is greater than the number of observations $n$). Group LASSO (Bakin $ 1999$; Cai $2001$, Antoniadis \& Fan $2001$;  Youan \& Lin $ 2006$  Meier et al. $2008$) handles these problems by extending the LASSO penalty to cover group variable structures.

Estimating coefficients in group LASSO is slightly different from standard LASSO because the constraints are now applied to each grouping of variables. In regular LASSO it is possible to have a different constraint for each coefficient. Group LASSO removes a set of explanatory variables in the model by shrinking its corresponding parameter to zero and keeping a subset of significant variables upon which the hazard function depends. As can be noticed, Group LASSO penalizes each factor in a very similar manner as  usual LASSO. In other words, same tuning parameter $ \lambda$ is used for each factor without assessing its relative importance. In a typical linear regression setting, it has been shown that such an excessive penalty applied to the relevant variables can degrade the estimation efficiency (Fan \& Li 2001) and affect the selection consistency (Leng et al. 2006; Yuan \& Lin 2006; Zou 2006). Therefore, it can reasonably be expected that Group LASSO suffers the same drawback. For linear regression problems, (Wang \& Leng 2008) proposed adaptive group LASSO, which allows for unique tuning parameter values to be used for separate factors. Such flexibility in turn produces different amounts of shrinkage for different factors. Intuitively, if a relatively large amount of shrinkage is applied to the zero coefficients and a relatively small amount is used for the nonzero coefficients, an estimator with a better efficiency can be obtained.

In the classic semi-parametric Cox model, the study population is implicitly assumed to be homogeneous, meaning all individuals have the same risk of death. This assumption rarely holds true. Individuals within a group may possess a non-observed susceptibility to death from differential genetic predisposition to certain diseases or have common environmental exposures that influence time to the studied event. Another standard assumption in the analysis of survival data is that the individuals under observation are independent. This assumption may be violated in many cases. We may observe a relationship among individuals of the same group when they share unobserved risk factors. Typical groups sharing some risk factors include families, villages, hospitals, and repeated measurements on one individual. A simple model for dependent survival times that is a generalization of the proportional hazard model can be implemented using the concept of $frailty$. This was first proposed by (Vaupel et al. $1979$).

The frailty distributions that have been studied mostly belong to the power variance function family, a particular set of distributions introduced first by Tweedy ($1984$) and later independently studied by Hougaard ($1986$). The gamma, inverse Gaussian, positive stable, and compound Poisson distributions are all members of this group. Generally, the gamma distribution is used to model frailty, mostly for mathematical convenience. It has been demonstrated that its Laplace transform is a useful mathematical tool for several measures of dependence, and the $ n^{th}$ derivative of its Laplace transform has a simple notation. To control the hidden heterogeneity and/or dependence among individuals with a group-related $"frailty"$, we introduce into our model a random variable that follows a gamma distribution. In frailty modeling, the gamma distribution is typically parametrized with one parameter being used simultaneously for both shape and scale.

In this context (Fan \& Li $2002$) proposed LASSO for the Cox proportional hazard frailty model. In this paper, we further improve this procedure by extending it to group LASSO for the Cox proportional hazard frailty model for survival censored times in high dimensions. Like classic LASSO, group LASSO shrinks and selects important predictors, taking into account group structure and known linkages between predictor variables that are supplied in the model. Additionally, allowance is made for a group-level frailty previously described that may be related to unmeasured but suspected background vulnerability or resilience to a particular disease outcome. This model algorithm, using group LASSO with the Cox proportional hazard frailty model, is most applicable in situations with the aforementioned characteristics. In this paper, we will provide a simulated situation and dataset that demonstrates how this method may be used.
\section{Methods}
\label{MaM}

\subsection{Model set-up} 

Suppose that there are n clusters and that the $i^{th}$ cluster has $J_i$ individuals and associates with unobserved shared frailty $u_i ( 1\leq i\leq n)$. A vector $ X_{ij} (1\leq i \leq n, 1 \leq j \leq J_i)$ is associated with the $ij^{th}$ survival time $T_{ij}$ of the $j^{th}$ individual in the $i^th$ cluster. Assume that  we have independent and identically distributed survival data for a subject $j$ in $i^{th}$ cluster: $(Z_{ij},\delta_{ij}, X_{ij}, u_i)$ with $ \delta_{ij}= \mathds{1}_{\lbrace T_{ij}\leq C_{ij} \rbrace}$ the status indicator of censoring, $C_{ij}$ the censoring time and $ Z_{ij} =min(T_{ij}, C_{ij})$ the observed time respectively for the individual $j$ of the cluster $i$. The corresponding likelihood function with a shared gamma frailty is given by:
\begin{equation}\label{eq:0}
L_n(\beta,H,\alpha)=\prod_{i=1}^{n} \prod_{j=1}^{J_i}\Bigg\{ h_{ij}\Big( Z_{ij}| u_i,X_{ij}\Big) ^{\delta_{ij}}S_{ij}\Big( Z_{ij}| u_i,X_{ij}\Big) \Bigg\} \prod_{i=1}^{n}g(u_i)
\end{equation}
with $S(t)=\exp(-H_0(t))$ a conditional survival function, $h(t| X,u)$ a conditional hazard function
of $T$ given $X$ and $ u$, and $$ g(u)=\frac{\alpha^\alpha u^{\alpha-1}\exp (-\alpha u)}{\Gamma(\alpha)}$$ the density function of a one-gamma frailty $u$ . 
Consider the  Cox proportional hazard with frailty model:
\begin{equation}\label{eq:1}  
h_{ij}(t| X_{ij}, u_{i}) = h_{o}(t)u_{i}\exp(\beta^{\top}X_{ij})
\end{equation}
with $h_o(t)$ the baseline hazard function and $\beta$ the parameter vector of interest, $ H_0(t)=\int^t_0 h_o(\mu)\,d\mu$ the cumulative baseline hazard function. Then (\ref{eq:0}) becomes:
\begin{equation}\label{eq:2}
\prod_{i=1}^{n} \prod_{j=1}^{J_i} h_0(Z_{ij})^{\delta_{ij}}\exp(\beta^{\top}X_{ij})u_i^{\delta_{ij}}\exp \lbrace-H_0(Z_{ij})\exp(\beta^{\top}X_{ij})u_i\rbrace \prod_{i=1}^{n}g(u_i).
\end{equation}

The likelihood of the observed  data is obtained by integrating (\ref{eq:2}) with respect to  $u_1,...,u_n$.\\ 

$$\int_{u_1}\dots\int_{u_n}\prod_{i=1}^{n} \prod_{j=1}^{J_i}\Big\{ h_0(Z_{ij})^{\delta_{ij}}\exp(\beta^{\top}X_{ij})u_i^{\delta_{ij}}\exp \Big [-H_0(Z_{ij})\exp(\beta^{\top}X_{ij})u_i\Big ] \Big \}\prod_{i=1}^{n}g(u_i)\,du_n\dots\,du_1$$\\

$$=\prod_{i=1}^{n} \prod_{j=1}^{J_i} h_0(Z_{ij})^{\delta_{ij}}\exp(\beta^{\top}X_{ij})*\underbrace{\int_{u_1}\dots\int_{u_n}\prod_{i=1}^{n}\Big \{ \prod_{j=1}^{J_i}u_i^{\delta_{ij}}\exp\Big [-H_0(Z_{ij})\exp(\beta^{\top}X_{ij})u_i \Big ]\Big \}\prod_{i=1}^{n}g(u_i)\,du_n\dots\,du_1}.$$\\

Let $A = \int_{u_1}\dots\int_{u_n}\Big \{ \prod_{i=1}^{n}u_i^{\sum_{j=1}^{J_i}\delta_{ij}}\exp \Big[-\sum_{j=1}^{J_i}H_0(Z_{ij})\exp(\beta^{\top}X_{ij})u_i\Big]\Big \} g(u_i)\,du_n\dots\,du_1$\\

$=\int_{u_1}u_1 ^{A_1}\exp\Big[-\sum_{j=1}^{J_1}H_0(Z_{1j})\exp(\beta^{\top}X_{1j})u_1\Big]\frac{1}{\Gamma(\alpha)}\alpha^\alpha u_1^{\alpha-1}\exp(-\alpha u_1)\,du_1 * \dots $ with the product continued for $ i=2,...n$ according to the format notated above for $i = 1$, with $ A_i=\sum_{j=1}^{J_i}\delta_{ij}$.\\

\begin{equation}\label{eq}             
L_n(\beta,\alpha,H_o)=\prod_{i=1}^n\underbrace{\int_{u_i}u_i ^{(A_i+\alpha)-1}\exp\Big\{-\Big[\sum_{j=1}^{J_i}H_0(Z_{ij})\exp(\beta^{\top}X_{ij})+\alpha\Big] u_i\Big\}\,du_i}*\prod_{i=1}^n\frac{\alpha^\alpha}{\Gamma(\alpha)}
\end{equation}
With a suitable change of variables,\\

$$L_n(\beta,\alpha,H_o)=\prod_{i=1}^n\Gamma(A_i+\alpha)\frac{1}{\Big[\sum_{j=1}^{J_i}H_0(Z_{ij})\exp(\beta^{\top}X_{ij})+\alpha\Big]^{A_i+\alpha}}\prod_{i=1}^n\frac{\alpha^\alpha}{\Gamma(\alpha)}$$ With $ A_i=\sum_{j=1}^{J_i}\delta_{ij}$
\begin{equation}\label{eq:3}
L_n(\beta,\alpha,H_o)=\prod_{i=1}^n \frac{\alpha^\alpha\prod_{j=1}^{J_i}h_0(Z_{ij})^{\delta_{ij}}\exp(\beta^{\top}X_{ij})\delta_{ij}}{\Gamma(\alpha)\Big[\sum_{j=1}^{J_i}H_0(Z_{ij})\exp(\beta^{\top}X_{ij})+\alpha\Big]^{A_i+\alpha}}\Gamma(A_i+\alpha)
\end{equation}\\
The logarithm of the likelihood in (\ref{eq:3}) is given by
\begin{equation}\label{equ:4}
\begin{split}
\ell_n(\beta, \alpha, H_0)&=\sum_{i=1}^n\Bigg\{ \alpha\log\alpha+\sum_{j=1}^{J_i}\big[\beta^{\top}X_{ij}\delta_{ij}+\delta_{ij}\log h_0(Z_{ij})\big]+\log \Gamma(A_i+\alpha)-\log\Gamma(\alpha)\\
& -(A_i+\alpha)\log\big[\sum_{j=1}^{J_i}H_0(Z_{ij})\exp(\beta^{\top}X_{ij})+\alpha\big]\Bigg\}
\end{split}
\end{equation}
\begin{equation}\label{eq:5}
\ell_n(\beta, \alpha, H_0) \equiv \sum_{i=1}^n \sum_{j=1}^{J_i} \delta_{ij}\log h_0(Z_{ij})-\sum_{i=1}^n(A_i+\alpha)\log\Bigg\{\sum_{j=1}^{J_i}H_0(Z_{ij})\exp(\beta^{\top}X_{ij})+\alpha\Bigg\}
\end{equation}
We formulate a profiled likelihood as follows: Consider the least informative nonparametric modeling for $H_0$ in which $H_0(Z)$ has a possible jump of size $\rho_l$ at the observed failure time $\tilde{Z_l}$. Then
\begin{equation}\label{eq:6}
\begin{split}
H_N(Z)&=\sum_{l=1}^N \rho_l \mathds{1}_{\lbrace\tilde{Z_l}\leq Z\rbrace}\\
h_N(Z_{ij})&=\prod_{l=1}^N \rho_l^{\mathds{1}_{\lbrace \tilde{Z_l} \leq Z_{ij}\rbrace}}
\end{split}
\end{equation} where $\tilde{Z_l},l=1,...,N$ are pooled observed failure times. Substituting  (\ref{eq:6}) in (\ref{eq:5}), we get:\\

\begin{equation}\label{eq:7}
\begin{split}
\ell_n(\beta, \alpha, H_N)& \equiv \sum_{i=1}^n \sum_{j=1}^{J_i} \delta_{ij}(\sum_{l=1}^N \mathds{1}_{\lbrace \tilde{Z_l}\leq Z_{ij}\rbrace}\log \rho_l)\\
&-\sum_{i=1}^n(A_i+\alpha)\log\Bigg\{\alpha+\sum_{j=1}^{J_i}\exp(\beta^{\top}X_{ij})\sum_{l=1}^N \rho_l \mathds{1}_{\lbrace \tilde{Z_l}\leq Z_{ij}\rbrace}\Bigg\}
\end{split}
\end{equation}\\

\begin{equation}
\begin{split}
\frac{\partial  \ell_n(\beta, \alpha, H_N) }{\partial \rho_k}&=\sum_{i=1}^n \sum_{j=1}^{J_i} \delta_{ij}\mathds{1}_{\lbrace \tilde{Z_k} \leq Z_{ij}\rbrace}\frac{1}{\rho_k}\\
&-\sum_{i=1}^n(A_i+\alpha)\frac{\sum_{j=1}^{J_i}\exp(\beta^{\top}X_{ij})\mathds{1}_{ \lbrace \tilde{Z_k}\leq Z_{ij}\rbrace}}{\alpha+\sum_{j=1}^{J_i}\exp(\beta^{\top}X_{ij})\sum_{l=1}^N \rho_l \mathds{1}_{\lbrace \tilde{Z_l}\leq Z_{ij}\rbrace}},    k=1,...N
\end{split}
\end{equation}\\

Assume there are no simultaneous events ("ties") occurring for different groups.
\begin{equation}\label{eq:9}
\frac{1}{\rho_k}=\sum_{i=1}^n\frac{(A_i+\alpha)\sum_{j=1}^{J_i}\exp(\beta^{\top}X_{ij})\mathds{1}_{\lbrace\tilde{Z_k}\leq Z_{ij}\rbrace}}{\alpha+\sum_{j=1}^{J_i}\exp(\beta^{\top}X_{ij})\sum_{l=1}^N \rho_l \mathds{1}_{\lbrace\tilde{Z_l}\leq Z_{ij}\rbrace}},  k=1,...N
\end{equation}
The value of  $\rho_k$ in  (\ref{eq:9}) is obtained numerically with the algorithm described section (4).
\subsection{Group LASSO estimator for Cox regression with frailty}

The objective function in the Group LASSO for Cox model with frailty  is
\begin{equation}\label{eq:9b}
Q_n(\beta, \lambda_n)=-\frac{1}{n}\ell_n(\alpha,\beta, H_N )+ \lambda_n\sum_{\substack{j=1}}^K \sqrt{p_j} \lVert \beta_{(j)} \rVert_2 
\end{equation}
where $Q_n(\beta, \lambda_n) $ is the objective convexe function to be minimized over the model parameter $\beta$ with a given optimal tuning parameter $\lambda_n$. This optimal turning parameter controls the amount of penalization. $\ell_n(\beta,\beta, H_N )$ is the profiled partial log-likelihood from (\ref{eq:7}).  The model parameter $\beta$ is decomposed into $K$ vectors $\beta_{(j)}, j=1,2,...,K$ which correspond to the $K$ covariate groups, respectively. The term $ \sqrt{p_j}$ adjusts for the varying group sizes, and  $ \lVert . \rVert_2$ is the Euclidean norm.

The group LASSO estimator for Cox regression with frailty is defined as: \\
\begin{equation}\label{eq:10}
\hat{\beta}_n(\lambda_n)=\arg\min_{\beta}\left\lbrace -\frac{1}{n}\ell_n(\alpha,\beta, H_N )+ \lambda_n\sum_{\substack{j=1}}^K \sqrt{p_j} \lVert \beta_{(j)} \rVert_2 \right\rbrace 
\end{equation}
This estimator does not have an explicit solution in general due to non-differentiability. Therefore, we use an iterative procedure to solve the minimization problem. Depending on the value of the optimal tuning parameter $ \lambda_n$, the estimated coefficients within a given parameter group $j$   satisfy: Either  $ (\hat{\beta}_{(j)}= 0)$  for all its components or   $ (\hat{\beta}_{(j)}\neq 0)$    for all its components. 
This occurs as a consequence of non-differentiability of the square root function at zero $ (\beta_{(j)} =0)$. If the group sizes are all one, the process reduces to the standard $LASSO $.

\subsection{Model selection - find an optimal tuning parameter $\lambda$}
It is necessary to have an automated method for selecting the tuning parameter $\lambda$  that controls the amount of penalization that is considered to be optimal dependent on a specific criterion, such as the Akaike information criterion (AIC) (Akaike, 1973), the Bayesian information criterion (BIC) (Schwarz 1978) or generalized cross-validation (GCV) (Craven and Wahba 1978). We would like to assign the best value to $\lambda$, however that is defined. There is no easy or universally agreed upon best way to find the optimal value for $\lambda$, or for any tuning parameter. In general, the selected value is based on optimizing some function, typically a loss function $\sum_{i=1}^n L(y_i ,\hat{f}(X_i))$ where $\hat{f}(X)$ is a prediction model fitted on a training subset of data. Finding the value for $\lambda$ that performs best according to the metric of choice can be done through several methods, of which k-fold cross-validation (CV) is the most common. In k-fold CV we randomly split the data into k so-called folds. For every fold $i = 1 ... k$, we fit a model on all available data less the data in that particular fold, which is used as the training set. With that model, we try to predict the data in the missing fold, known as the test set. For each fold we obtain an estimate of some metric to evaluate our model, such as an evaluation of a relevant loss function. As a final estimate of how our model performs, we take the average metric over all of the folds. The cross validation error for the subset is naturally chosen to be the negative log likelihood. An important problem of k-fold CV is the computational burden. Fitting a penalized proportional hazards model is computationally intensive, especially if the model has to be fit multiple times for each value of $\lambda$ we want to evaluate. In this paper, choosing $k$ to be equal to $10$, we estimate $\lambda$ by minimizing a k-Cross Validation( GCV)  error that is mathematically illustrated as follows: $$ CV_k(\lambda)=-\sum_{i=1}^{k}\ell_n^i\Big(\hat{\beta}_{(n-i)}(\lambda)\Big)/n$$ $\hat{\beta}_{(n-i)}(\lambda)$ is the penalized estimate for $ \beta$ at $ \lambda$ with the $ i^{th}$ subset taken out as the test set and the remaining $k-1$ subsets kept as the training set. $ \ell_n^i(.)$ is the log partial likelihood for the $ i^{th}$ subset.


\section{ Algorithm}\label{sec:1}
To minimize (\ref{eq:9b}) we use the following procedure: We split (\ref{equ:4}) into two pseudo log-likelihood functions. One mainly depending on $\beta$ :
\begin{equation}\label{eq:12}
\ell_n^{(\beta)}(\beta, \alpha, H_N)\equiv\sum_{i=1}^n\sum_{j=1}^{J_i}\beta^{\top}X_{ij}\delta_{ij}-\sum_{i=1}^n(A_i+\alpha)\log\Bigg\{\sum_{j=1}^{J_i}H_N(Z_{ij})\exp(\beta^{\top}X_{ij})+\alpha\Bigg\}
\end{equation}  
and the other mainly depending on $ \alpha$:
\begin{equation}\label{eq:13}
\ell_n^{(\alpha)}(\beta,\alpha,H_N)\equiv\sum_{i=1}^n\Bigg\{\alpha\log\alpha+\log \Gamma(A_i+\alpha)-\log\Gamma(\alpha)-(A_i+\alpha)\log\Big[\sum_{j=1}^{J_i}H_N(Z_{ij})\exp(\beta^\top X_{ij})+\alpha\Big]\Bigg\}
\end{equation}
Since the the penalty term in (\ref{eq:9b}) depends only on $ \beta$, minimizing (\ref{eq:9b}) is equivalent with minimizing:
\begin{equation}\label{eq:14}
-\frac{1}{n}\ell_n^{(\beta)}(\beta, \alpha, H_N)+ \lambda_n\sum_{\substack{j=1}}^K \sqrt{p_j} \lVert \beta_{(j)} \rVert_2 
\end{equation} 

We cycle through the parameter groups and minimize (\ref{eq:14}) keeping all except the current parameter group fixed. The Block Co-ordinate Gradient Descent algorithm is to be applied to solve the non-smooth convex optimization problem in (\ref{eq:14}) (Yun et al. 2011). This algorithm would also be used to optimize (\ref{eq:13}). However, (\ref{eq:13}) involves the first two order derivatives 
of the gamma function, which may not exist for certain values of $\alpha$. We use an approach similar to that in (Fan \& Li 2002) to avoid this difficulty by using a grid of possible values for the frailty parameter $ \alpha$ and finding the minima of (\ref{eq:13}) over this discrete grid, as suggested by Nielsen et al. (1992).

Denote  $Q_{\lambda_n}(\beta)= -\frac{1}{n}\ell_n^{(\beta)}(\beta, \alpha, H_N)+ \lambda_n\sum_{\substack{j=1}}^K \sqrt{p_j} \lVert \beta_{(j)} \rVert_2 $ a penalized objective function  to be minimized and denote $ \nabla Q_{\lambda_n}(\beta)$ its gradient to be evaluated at $\beta$\\

\begin {table}[!ht]
\caption {Block Co-ordinate Gradient (BCGD) Descent Algorithm} \label{tab:title} 
\begin{tabular}{l|p{10cm}}\hline
	Steps  & Algorithm \\\hline
	1.   & For $j= 1,...,K$ \newline choose $ \hat{\beta}_{(j)}^{(0)}$ as initial values.\\
	2.   & For the $m^{th}$ iteration, $ \hat{\beta}_{(j)}^{(m+1)} \leftarrow \hat{\beta}_l^{(m)}-\gamma_n \nabla Q_{\lambda_n}(\hat{\beta}_l^{(m+1)}) $ with  $ m =0,1,2,... $ and $ \gamma_n>0$ the step size computed following Armijo rule \\ 
	3.&  For each $j$, repeat steps 2 until some convergence criterion is met\\ \hline
	
\end{tabular}
\end {table}

With BCGD, we propose the following algorithm to solve (\ref{eq:9b}).

\begin{tabular}{l|p{10cm}}\hline
	Steps  & Algorithm \\\hline
	1.   & For $j= 1,...,K$ \newline choose $ \hat{\beta}_{(j)}^{(0)},\hat{\alpha}_{(j)}^{(0)}, \hat{\rho}_{j,k}^{(0)}$, k=1,...,N  as initial values.\\
	2.   & For the $m^{th}$ iteration, $\hat{\rho}_{j,k}^{(m+1)}$ is updated from (\ref{eq:9} ) with $ m =0,1,2,... $ and then compute $\hat{H}_{N}^{(m+1)}$ from (\ref{eq:6})\\ 
	3.   & Since $\hat{H}_{N}^{(m+1)}$ is known, we can then minimize (\ref{eq:13}) with respect to $\left( \hat{\beta}_{(j)}^{(m+1)}\right)$  using BCGD algorithm\\
	4.& Since $\left(\hat{H}_{N}^{(m+1)} ,\hat{\beta}_{(j)}^{(m+1)}\right)$ are known, we minimize (\ref{eq:14}) with respect to $\left( \hat{\alpha}_{(j)}^{(m+1)}\right) $ as stated above\\
	5.&  For each $j$, repeat steps 2 up 4 until some convergence criterion is met\\ \hline
	
\end{tabular}\\

\section{Theoretical consistency of the method}

Consider the penalized pseudo-partial likelihood estimator: $$  \hat{\beta}_n(\lambda_n)=\arg\min_{\beta}\left\lbrace -\frac{1}{n}\ell_n(\alpha,\beta, H_N )+ \lambda_n \sum_{\substack{j=1}}^K \sqrt{p_j} \lVert \beta_{(j)} \rVert_2 \right\rbrace $$ Denote $\beta^0 $ the true value of the model parameter $\beta=(\alpha,\beta, H_N)$. $ \forall \varepsilon > 0$, we need to show that $\mathds{P}\left\lbrace \hat{\beta_n}(\lambda_n)-\beta^0\|<\varepsilon \right\rbrace\rightarrow 1$ as $ n \to \infty$. Given (A)-(D) regularity conditions in (Andersen and Gill $1982$), according to the Theorem 3.2 in Andersen and Gill (1982), the following two results hold.
$$ n^{-1/2}\dot{\ell}_n(\beta^0)\overset{\mathds{P}}{\to}\mathcal{N}(0, \Sigma)$$
$$ -\frac{1}{n}\ddot{\ell}_n(\beta^\ast)\overset{\mathds{P}}{\to}\Sigma \hspace{3mm} \forall \hspace{2mm}\beta^\ast\overset{\mathds{P}}{\to}\beta^0$$ $\dot{\ell}_n(\beta^0)$ and $ \ddot{\ell}_n(\beta^\ast)$ are the first and the second order derivatives of $\ell_n(\beta)$, i.e, the score function and the Hessian matrix, evaluated at $\beta^0$ and $\beta^\ast$ respectively. $\Sigma$ is the positive definite Fisher information. The consistency theorem stated in this section buids up on the two results above.
\pagebreak
\begin{Theorem}(Consistency)
	Assume that $(X_{ij},T_{ij},C_{ij})$ are independently distributed random samples given $u_i$ which are i.i.d. from a Gamma distribution for $i=1,...,n$ and $j=1,...,J_i$. $ T_{ij}$ and $ C_{ij}$ are conditionally independent given $X_{ij}$. Under regularity conditions (A)-(D) in Anderson and Gill (1982), if $\lambda_n\to 0$ when $ n \to \infty$, then there exists a local minimizer $\hat{\beta_n}(\lambda_n)$ of $ Q_n(\beta, \lambda_n)$ such that $ \mathds{P}\left\lbrace \| \hat{\beta_n}(\lambda_n)-\beta^0\|<\varepsilon\right\rbrace\rightarrow 1$
\end{Theorem}

Proof: Applying Theorem 5.7 in Van der Vaart(1998) with a slightly different approach the theorem can be proved as follows: Let us first show that $ Q_n(\beta_n, \lambda_n)>Q_n(\beta_n^0, \lambda_n).$
$$ Q_n(\beta_n, \lambda_n)-Q_n(\beta_n^0, \lambda_n).$$
$$ =-\frac{1}{n}\left( \ell_n(\beta)-\ell_n(\beta^0)\right)+\sum_{j=1}^K \lambda_n \sqrt{p_j}\left( \| \beta_{(j)}\|-\| \beta^0_{(j)}\|\right)  $$
$$ \geq-n^{-1/2}\left(n^{-1/2}\frac{\partial}{\partial\beta}\Big(\ell_n(\beta^0)\Big) \right)^\top\left(\beta-\beta^0 \right)+ \left(\beta-\beta^0 \right)^\top \left( n^{-1/2}\frac{\partial^2}{\partial\beta^2}\Big(\ell_n(\beta^0)\Big)\right) \left(\beta-\beta^0 \right)$$ $$+n^{-1}o_p\left( \|\beta- \beta^0\|^2\right)-\sum_{j=1}^K \lambda_n \sqrt{p_j}\left( \| \beta_{(j)}\|-\| \beta^0_{(j)}\|\right)  $$
$$ \geq-n^{-1}O_p(1) \|\beta- \beta^0\|+\left(\beta-\beta^0 \right)^\top\left( \Sigma+o_p(1)\right)\left(\beta-\beta^0 \right)+n^{-1}o_p\left( \|\beta- \beta^0\|^2\right) -\lambda_n \sum_{j=1}^K \sqrt{p_j}\left( \| \beta_{(j)}\|-\| \beta^0_{(j)}\|\right) $$ Since $\lambda_n\rightarrow 0$ as $n\rightarrow0$ then $Q_n(\beta_n, \lambda_n)-Q_n(\beta_n^0, \lambda_n)\geq \left(\beta-\beta^0 \right)^\top\left( \Sigma+o_p(1)\right)\left(\beta-\beta^0 \right)$ and this right side part is positive since $\Sigma$ is positive. $Q_n(\beta_n, \lambda_n)$ is non empty and lower bounded by $Q_n(\beta_n^0, \lambda_n)$ consequently it admits a local minimum. Since $Q_n(\beta_n, \lambda_n)$ is concave, its local minimum is also its global minimum.
$$ Q_n(\beta_n, \lambda_n)>Q_n(\beta_n^0, \lambda_n).$$For any positive $\varepsilon$ 
$$ \left\lbrace \sup_{\beta:\|\beta- \beta^0\|=a}Q_n(\beta_n, \lambda_n)>Q_n(\beta_n^0, \lambda_n)\right\rbrace  \subseteq \left\lbrace \hat{\beta_n}(\lambda_n)-\beta^0\|<\varepsilon\right\rbrace $$
$$ \Rightarrow \mathds{P}\left\lbrace \hat{\beta_n}(\lambda_n)-\beta^0\|<\varepsilon\right\rbrace \geq \mathds{P}\left\lbrace \sup_{\beta:\|\beta- \beta^0\|=a}Q_n(\beta_n, \lambda_n)>Q_n(\beta_n^0, \lambda_n)\right\rbrace $$
Thus $ \Rightarrow \mathds{P}\left\lbrace \hat{\beta_n}(\lambda_n)-\beta^0\|<\varepsilon\right\rbrace\rightarrow 1$

\section{ Applications}

With the advent of molecular biology to study the relationship between genetics and disease outcomes such as cancer, and as exposure science improves for taking multiple polluant or pathogen measurements, in air and water as well as in other media, it becomes possible for affected individuals, researchers and public health practitioners to generate large datasets with rich information such that the numbers of predictors $p$ is greater than the sample sizes $n$. Statistical methods are needed to handle and analyze such data sets. In the case of genetic epidemiology, researchers are able to identify genes that act along identical or similar pathways and are able to group these genes together to understand associations with health outcomes and to calculate cumulative risk. In the case of exposure assessment, environmental health scientists now understand that pollution sources release multiple pollutants that contribute to the same morbidities. Examples include the many chemicals in tobacco smoke, vehicle emissions, and effluents from industrial plants. People experiencing diarrhea may have co-infection with multiple pathogenic agents, and understanding the nature of outbreaks may be improved as water exposure science advances in the future. Personalized medicine has opened the door to personalized public health as more information can be gathered at the individual level. By using group LASSO with group level frailty in survival analysis, we will be better able to trace health outcomes back to sources that contribute multiple exposures of interest. Group LASSO's preferential shrinking towards zero of non-significant groups of predictors will produce sparse models that link back to pollution sources rather than individual chemical or biological exposures. This application could be applied in the case of land-use studies, brownfield risk assessment, and environmental impact assessments of new construction projects. Group LASSO with the Cox proportional hazards frailty model will be part of the new paradigm of risk assessment that encompasses cumulative exposures (National Research Council of the National Academies 2009). For use with genetic epidemiology, as gene mapping and gene testing become increasingly cost effective, large cohort datasets will become available to more effectively establish associations between genetic and epigenetic markers and disease outcomes. As previously discussed, group LASSO with group frailty allows common pathways and mechanisms to be incorporated into the analysis while also including a frailty term to account for unmeasured susceptibility or resilience that exist in subpopulations.

\section{ Simulated data}
Data sets were simulated with sample size $m = \sum_{i=1}^{n}J_i$ (where $n$ is the number of observation clusters and $J_i$ is the number of observations in the $i^{th}$ cluster) fixed to $100$ and predictors $p$ equals to $100$. Group sizes for both individuals (with respect to frailty) and predictors (with respect to variable groupings) were set to 10 arbitrarily, though this can easily be adjusted depending on the dataset. We simulated a design matrix of of order $(n,p)$ where $X_i\overset{i.i.d}{\to}\mathcal{N}(0,1)$ and the covariance matrix $\Sigma_{i,j}=\rho^{| i-j|}$ with $\rho=0.5$. In practice, the assumption of a constant hazard function is rarely tenable. A more general form of the hazard function is given by the Weibull distribution, which is characterized by two positive parameters: the scale parameter$(\lambda>0)$ and the shape parameter ($\nu >0$). Its corresponding baseline hazard function is $$ h_0(t)=\lambda\nu t^{\nu-1}$$ and the survival time for a shared-gamma frailty Cox model is $$ T=\left( \frac{\log(U)\exp(-\beta^\top X)}{\lambda G}\right)$$ with $ U\leadsto Uni[0,1]$ and $G\leadsto\Gamma(\alpha,\alpha)$. Taking into account the censoring status, we simulated censoring times from the exponential distribution: $C\leadsto\exp(n,3)$. The observed failure time for each observation is the minimum between its survival time $T$ and and its censoring status $C$.
The algorithms described in (\ref{sec:1}) were implemented to select the appropriate tuning parameter $\lambda$ to maximize the k-fold CV criterion. Performance of group LASSO with Cox proportional hazard frailty model is compared and contrasted with group SCAD and group MCP. Figures 1-3 show an example solution path for group LASSO, group SCAD, and group MCP, respectively.

\begin{figure}[h!]
	\includegraphics[width = 4in]{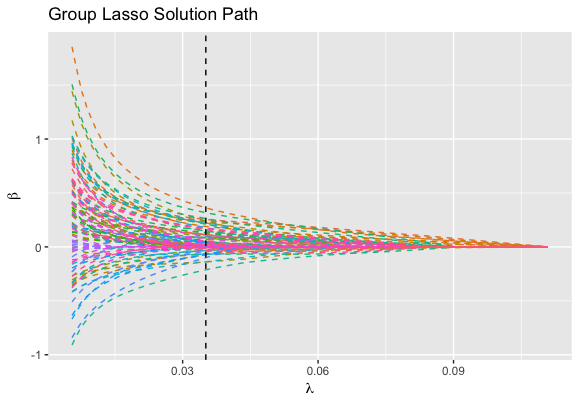}
	\centering
	\caption{Group Lasso Solution path for simulated examples}
	\label{fig:Sln path Lasso}
\end{figure}
\begin{figure}[h!]
	\includegraphics[width=4in]{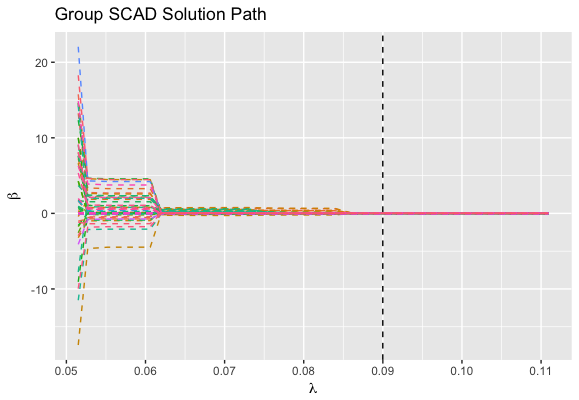}
	\centering
	\caption{Group SCAD Solution path for simulated examples}
	\label{fig:Sln path SCAD}
\end{figure}
\begin{figure}[h!]
	\includegraphics[width=4in]{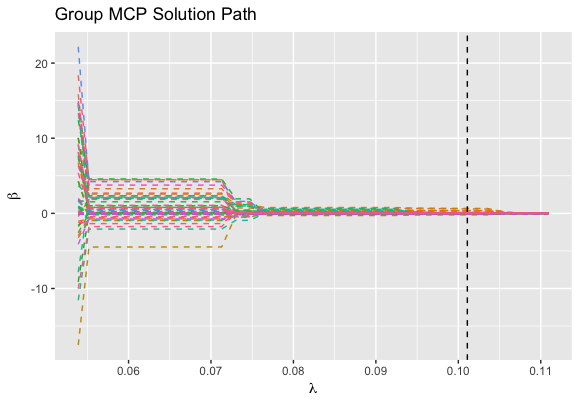}
	\centering
	\caption{Group MCP Solution path for simulated examples}
	\label{fig:Sln path MCP}
\end{figure}

\pagebreak

Figures 4-6 compare the performance of the three methods over 100 simulations with summary measures of tuning parameter value choice, cross-validation error, and R-squared, respectively (remembering that this is a simulated data set). Some summary trends appear. Notably for these simulations, group lasso tends to pick a smaller tuning parameter value, centered around 0.03 compared with 0.09 for group SCAD and 0.10 for group MCP. R-squared performance for group lasso is significantly better, averaging around 0.18 compared with 0.05 for group SCAD and 0.03 for group MCP. Considering cross-validation error, the results are more similar, with group lasso demonstrating only slightly better performance (139 for group lasso compared with 151 for group SCAD and 156 for group MCP) in this set of simulations.

\begin{figure}[h!]
	\includegraphics[width=4in]{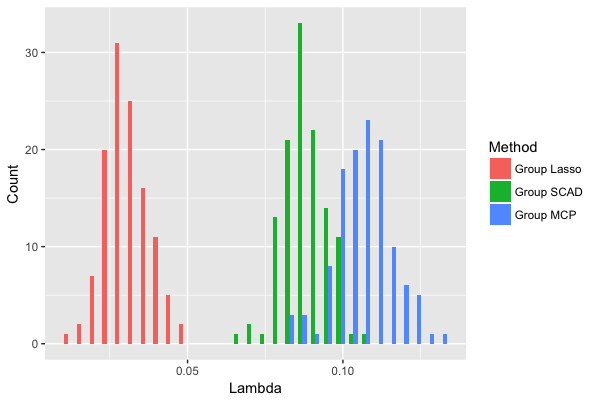}
	\centering
	\caption{Distribution of tuning parameter for each of the three methods over 100 simulations.}
	\label{fig:Compare Tuning Parameter}
\end{figure}

\begin{figure}[h!]
	\includegraphics[width=4in]{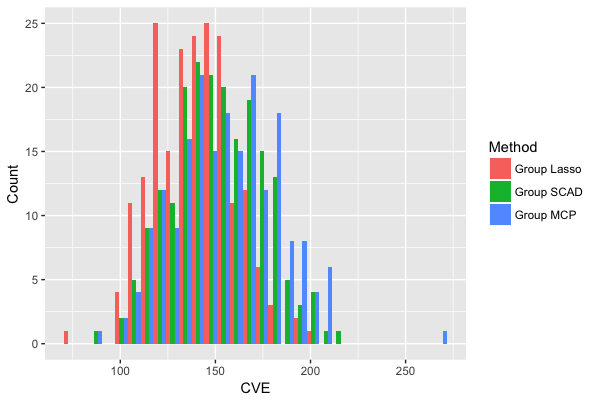}
	\centering
	\caption{Distribution of cross-validation errors for each of the three methods over 100 simulations.}
	\label{fig:Compare CVE}
\end{figure}

\begin{figure}[h!]
	\includegraphics[width=4in]{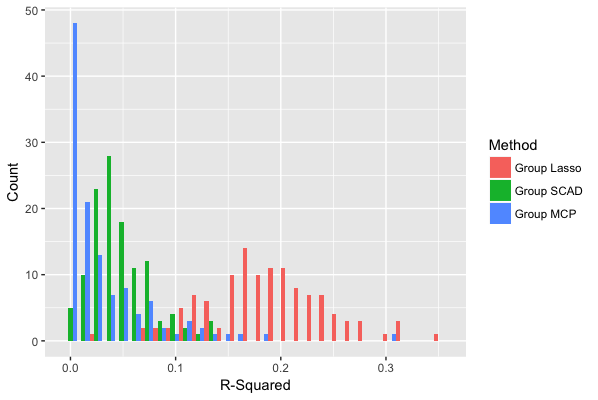}
	\centering
	\caption{Distribution of R-squared values for each of the three methods over 100 simulations.}
	\label{fig:Compare R-squared}
\end{figure}

\section{ Discussion }

The limitations of this methodology overlap with the limitations of LASSO. Group LASSO remains a penalization method that is not appropriate for all studies and circumstances and is outperformed at times by ridge regression, least-angle regression (LARS), and the non-negative garrrotte (Yuan and Lin $2007$). Even though group LASSO and group frailty make adjustments to account for clustering effects, this method requires a resolution of data and background knowledge that is not available for many data sets and research questions. Future research will continue to elucidate many of these scenarios and make the datasets more amenable to use with group LASSO. 
While the group LASSO gives a sparse set of groups, if it includes a group in the model then all coefficients in the group will be nonzero. Sometimes we would like parsimony both between groups and within each group. As an example, if the predictors are genes, we would then like to identify particularly "important" genes in the pathways of interest. Toward this end (Friedman et al. 2010) focused on the "sparse-group LASSO" wherein they introduced a regularized model for linear regression with $L_1$ and $L_2$ penalties. They discussed the sparsity and other regularization properties of the optimal fit for this model and show that it has the desired effect of group-wise and within group sparsity. Even though the group LASSO is an attractive method for variable selection, since it respects the grouping structure in the data, it is generally not selection consistent and also can select groups that are not important in the model (Wei and Huang $2011$). To improve the selection results, researchers proposed an adaptive group LASSO method which is a generalization of the adaptive LASSO and requires an initial estimator. They showed that the adaptive group LASSO is consistent in group selection under certain conditions if the group LASSO is used as the initial estimator. 
In this context, interested researchers may look into the "sparse-group LASSO" or "adaptive group LASSO" for use with the Cox proportional hazard model with frailty when optimizing grouped variable selection.

\section*{ Acknowledgments}
Funding for the initial meeting of authors JCU, TML, and PN was provided through MMED -  the Center for Inference and Dynamics of Infectious Diseases and funding provided through MIDAS-National Institute of General Medical Sciences under award U54GM111274.





\pagebreak

\section*{References}



\pagebreak





\pagebreak




\end{document}